# HABITABLE PLANETS ECLIPSING BROWN DWARFS: STRATEGIES FOR DETECTION AND CHARACTERIZATION

Adrian R. Belu[1,2], Franck Selsis[1,2], Sean N. Raymond[1,2], Enric Pallé[3,4], Rachel Street[5],
D. K. Sahu[6], Kaspar von Braun[7], Emeline Bolmont[1,2], Pedro Figueira[8], G. C. Anupama[6],
Ignasi Ribas[9]

[1] Univ. Bordeaux, LAB, UMR 5804, F-33270, Floirac, France.
[2] CNRS, LAB, UMR 5804, F-33270, Floirac, France
[3] Instituto de Astrofísica de Canarias, La Laguna, E38205 Spain
[4] Departamento de Astrofísica, Universidad de La Laguna, Av., Astrofísico Francisco Sánchez, s/n E38206-La Laguna, Spain
[5] Las Cumbres Observatory Global Telescope Network, 6740 Cortona Drive, Suite 102, Goleta, CA 93117, USA
[6] Indian Institute of Astrophysics, Koramangala, Bangalore 560034, India
[7] NASA Exoplanet Science Institute, California Institute of Technology, MC 100-22, Pasadena, CA 91125, USA
[8] Centro de Astrofísica, Universidade do Porto, Rua das Estrelas, 4150-762 Porto, Portugal
[9] Institut de Ciències de l'Espai (CSIC-IEEC), Campus UAB, Facultat de Ciències, Torre C5, parell, 2a pl., 08193 Bellaterra, Spain

## ABSTRACT

Given the very close proximity of their habitable zones, brown dwarfs represent high-value targets in the search for nearby transiting habitable planets that may be suitable for follow-up occultation spectroscopy. In this paper we develop search strategies to find habitable planets transiting brown dwarfs depending on their maximum habitable orbital period ($P_{HZ\,out}$). Habitable planets with $P_{HZ\,out}$ shorter than the useful duration of a night (e.g. 8-10 hrs) can be screened with 100% completeness from a single location and in a single night (near-IR). More luminous brown dwarfs require continuous monitoring for longer duration, e.g. from space or from a longitude-distributed network (one test scheduling achieved - 3 telescopes, 13.5 contiguous hours). Using a simulated survey of the 21 closest known brown dwarfs (within 7 pc) we find that the probability of detecting at least one transiting habitable planet is between $4.5^{+5.6}_{-1.4}$ and $56^{+31}_{-13}$ %, depending on our assumptions. We calculate that brown dwarfs within 5-10 pc are characterizable for potential biosignatures with a 6.5 m space telescope using ~1% of a 5-year mission's lifetime spread over a contiguous segment only $1/5^{th}$ to $1/10^{th}$ of this duration.

## 1. INTRODUCTION

Together with in-situ robotic exploration within our solar system, observation of terrestrial extra-solar planets' spectra are the current most robust approaches in the search for non-Earth life. The thermal emission from a habitable planet at 10 pc is ~1 photon sec$^{-1}$ m$^{-2}$ µm$^{-1}$; recording such a spectrum is within the capabilities of upcoming and even some existing space telescopes. Free-floating (rogue) planets, if yielding sufficient internal heat flow, could maintain habitable surface conditions if they also have adequate insulation, but this insulation would then limit the levels of photon emission, enabling characterization only for very nearby objects (1,000 AU for the case of solid insulation, Abbot & Switzer 2011). Therefore the main approach until now has been to search for characterizable habitable planets around beacon-primaries. These beacon-primaries then become the dominant noise source when subsequently undertaking the characterization of the exoplanets.

Resolving the planet from the primary is therefore the first challenge. The spatial resolution of planets remains a technological challenge (Traub et al. 2007, Cockell et al. 2009). Fortuitously, transiting planets[1] can be *time-resolved* from the brighter primary. Differential eclipse spectroscopy has enabled the identification of molecules in the atmospheres of giant planets close to solar-type stars (Tinetti et al. 2007; Grillmair et al. 2008; Swain et al. 2009; Stevenson et al. 2010, Beaulieu et al. 2010). However, even the upcoming James Webb *Space Telescope* (*JWST*) will be able to detect biomarkers with this technique only up to ~10 pc and for primary dwarves approximately M5 and later (Beckwith 2008, Kaltenegger & Traub 2009, Deming et al. 2009, Belu et al. 2011, Rauer et al. 2011, Pallé et al. 2011). Projects such as MEARTH (Charbonneau et al. 2008) are currently screening the solar neighborhood for these eclipsing planets.

---
[1] The planet's orbit is passing in front of the star as seen from the telescope (primary eclipse)

Yet the geometric transit likelihood and the local stellar population density & distribution do not guarantee the presence of even a single nearby transiting habitable planet suitable for characterization with *JWST* (Belu et al. 2011).

To help solve this scarcity problem we propose to extend the search to primaries not yet considered by current surveys: brown dwarfs (BDs). Bright primary objects (*primaries* hereafter) are indeed required for transit spectroscopy, but as we quantify in this paper, occultation (secondary eclipse) spectroscopy (in emission) of habitable planets around a nearby BD *is* a favorable scenario. The intrinsic emission of a body which is at given (habitable) equilibrium temperature is independent from the type of primary. On the contrary: the dimmer the primary, the less photon noise added to the planetary photons.

As said primaries are convenient 'signposts', 'lighthouses' or 'beacons' for planets, but once the planets are found the primaries become a barrier to planetary characterization. So what is the dimmest lighthouse? Jupiter-sized objects colder than room temperature have been detected with the *Wide-field Infrared Survey Explorer - WISE* (Cushing et al. 2011). Brown dwarf primaries therefore should represent the optimal limit for the 'lighthouse' search paradigm for habitable exoplanets that can be time-resolved[2].

But can planets form and remain habitable around BDs? There are significant differences between the potential habitability of planets around BDs and main sequence stars. For instance, BDs cool in time and their habitable zone (HZ) moves inward such that a planet on a stationary orbit sees the HZ sweep by in a much shorter time interval than for stellar dwarfs (Caballero & Rebolo 2002, Andreeschev & Scalo 2004). However, there is no clear "Achilles' heel" that would rule out BDs as habitable planet hosts; this issue is discussed at length in § 6. We thus continue with the assumption that BDs can indeed host habitable planets.

This paper is structured as follows. We first examine the observational characteristics of habitable planets eclipsing a BD (§ 2), which leads us to draft specific strategies for different regions of the BD parameter space (§ 3). We also find it useful (while keeping in mind the uncertainties on such estimates) to derive the contribution of the population of BD primaries to the expected number of transiting habitable planets sufficiently close to be characterized through occultation spectroscopy, regardless of the type of their primary (§ 4). We examine the performance of occultation spectroscopic characterization for a habitable planet eclipsing a BD over the whole BD parameter space (§ 5). We then review the literature on formation of terrestrial planets in the HZ of BDs, and discuss BD HZs. Finally in the Conclusion we outline the key points relevant to the fast-track roadmap of time-resolving characterization of habitable exoplanets.

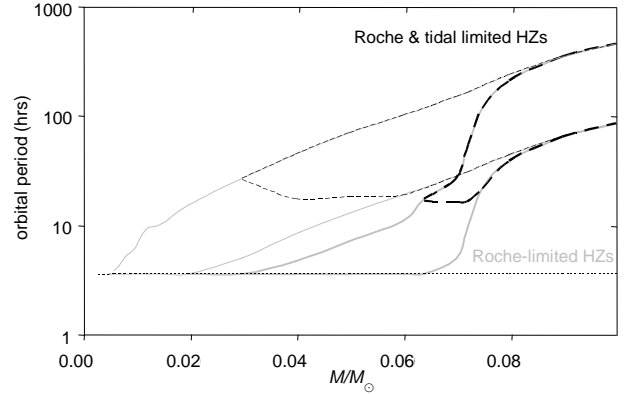

**Figure 1.** Orbital periods (in hours) of the effective habitable zone limits for 1 (**thin lines**) and 10 (**thick lines**) Gyr-old BDs. The restricted black zones account for depletion of the inner HZs by tidal migration (see text). The dotted line corresponds to the Roche period limit for the 10 $M_\oplus$, 1.8 $R_\oplus$ super-Earth considered here: 3.6 h.

## 2. OBSERVATIONAL CHARACTERISTICS OF HABITABLE PLANETS ECLIPSING A BROWN DWARF

The detectability of eclipsing habitable planets around BDs has already been considered & attempted (Caballero & Rebolo 2002, Caballero 2010): because of the small radius of the BD, putative terrestrial planets cause 1-5% transit depths. Blake et al. (2008, hereafter BL08) additionally note that the small orbital radius of a habitable planet around a BD (a few times $10^{-3}$ AU) increases its likelihood to transit. They also address issues such as the convenient ruling-out of background blend false positives thanks to the high proper motion of a nearby BDs population.

Figure 1 shows the orbital periods (in hours) of the effective habitable zones. We must caution here that the formula used for computing the habitable zone (Selsis et al. 2007a) is in principle valid for photospheric temperatures down to 3,700 K (fit to models of the Earth around F-G-K dwarfs). For the lower photospheric temperatures of M dwarfs in Belu et al. (2011) the correction for photospheric temperature was fixed at the 3,700 K value, and we do the same here: the scaling of the HZ is performed only through the luminosity of the primary (discussion of BD HZ in § 6).

Ford et al. (2006) show that if the planets undergo circularization then the limit of the possible orbital distances is twice the Roche limit, but if the planets undergo migration (such as in a disk) with a circular orbit, they can have orbital distances up to the Roche limit. We therefore also overplot in Fig. 1 the orbital period at the Roche limit $P_{\rm Roche}$ using the gravitation-only formula for the Roche $a_{\rm Roche}$ limit used in BL08 (citing Faber et al. 2005 and Paczyński 1971). Since $a_{\rm Roche} \propto M_{\rm BD}^{1/3}$ (the BD's mass),

---
[2] So, put in another way, we simply attempt here to push the astrobiological "follow the (liquid) water" philosophy to one of its many limits.

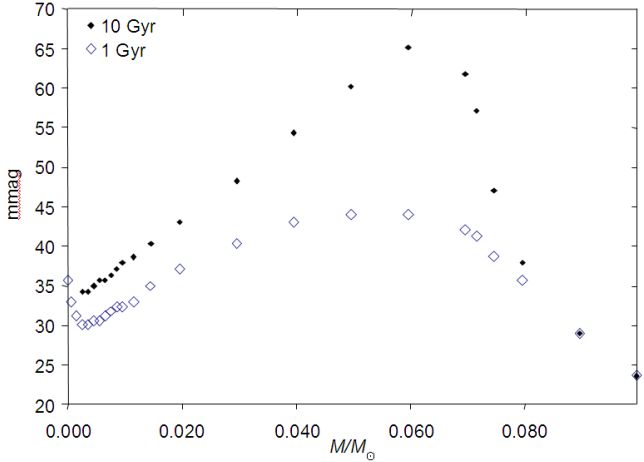

**Figure 2.** Transit depth of a 1.8 $R_⊕$ planet as function of the brown dwarf's (BD) mass for different ages of the BD.

$$P_{Roche} = \sqrt{4\pi \frac{a_{Roche}^3}{G(M_{BD}+M_P)}} \quad (1)$$

is fairly constant with the BDs mass, because the terrestrial planet's mass $M_P \ll M_{BD}$. $G$ is the gravitational constant. For all tidal migration aspects (here and hereafter): see § 4.2 hereafter.

While photometrically monitoring SIMP J013656.5+093347 for intrinsic variability, Artigau et al. (2009) detect a 50 mmag deep transit-like event in $J$ band, with a precision of 5 mmag over 5 min bins (1.6 m telescope at *Observatoire du Mont Mégantique*). The simultaneous monitoring in another band yielded a different depth of the event, which, they conclude, would not be the case if an opaque body such as a planet were masking the BD (the signature of a transit is grey). Therefore transit detection around a BD ideally involves simultaneously monitoring in two different bands.

For reference, Fig. 2 shows the depth in mmag of the transit of a 1.8 $R_⊕$ planet as function of the BD mass for different ages of the BD, with BD radii values from Baraffe et al. (2003, COND03 model).

Brown dwarfs are also fast rotators. As they contract and cool down on sub-Gyr timescales their rotation speeds increase. . We therefore note here that if the BD is significantly oblate and the planet's orbit is aligned with the BD's spin, this could reduce the transit depth because of equatorial gravity darkening (mention of this phenomenon in Herbst et al. 2007). For BDs older than 1 Gyr and heavier than 0.04 $M_⊙$ Bolmont et al. (2011 – Fig. 1) predict rotation periods below 1 hour (for reference, the rotation period of Jupiter is ~10 hr and its flattening 0.06). However, the authors recognize that measured BD rotational velocities available to fit their model are scarce, so caution is appropriate for this matter[3].

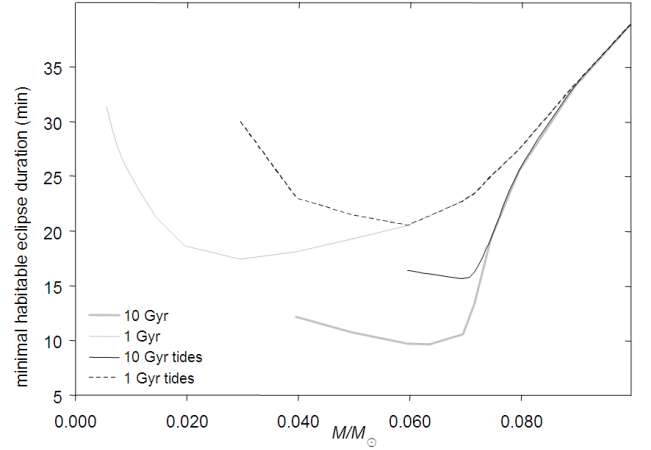

**Figure 3.** Minimal habitable planet eclipse duration (in minutes), as function of the brown dwarf's (BD) mass, for different ages of the BD, without and with tidal migration (optimistic planet formation at 10 Myr considered). For masses below the plotted low mass bounds the radiative habitable zone is entirely below the Roche limit or below the minimal asymptotic final tidal-migration orbit (no effective planetary habitable zone).

In the case of unresolvable binary BDs (such as 2MASS 0939-2448 AB, Leggett et al. 2009), an Earth-like planet around the brightest component (S-type orbit) still produces a 3.5% (40 mmag) deep transit in the combined photometry. For other unresolvable binary BDs, if both feature an effective HZ, one survey actually can monitor two habitable zones simultaneously.[4]

The required cadence of observation is set by the expected minimal duration of a habitable planet eclipse around a BD (Fig. 3):

$$\tau_{min} = P_{HZmin} \frac{1}{\pi} \operatorname{asin} \frac{\sqrt{1-b^2} R_{BD}}{\max(a_{HZin}; a_{Roche}; a_{tides})} \quad (2)$$

where $P_{HZ\,min} = \max(P_{HZ\,in}; P_{Roche}; P_{tides})$ is the effective minimal habitable circular orbital period. $P_{HZ\,in}$ is the orbital period at the inner limit $a_{HZ\,in}$ of the radiative habitable zone (HZ in), which is established for a planetary body. $P_{Roche}$, $P_{tides}$ and $a_{Roche}$, $a_{tides}$ are defined in the same way, at the Roche limit and for tidal migration respectively. $R_{BD}$ is the BDs radius, and $b$ is the median impact factor:

$$b = \frac{\max(a_{Roche}; a_{HZin})}{R_{BD}+1.8R_⊕} \cos\left[\frac{\pi}{4} + \frac{1}{2}\operatorname{acos} \frac{R_{BD}+1.8R_⊕}{\max(a_{Roche}; a_{HZin})}\right] \quad (3)$$

for a 10 $M_⊕$, 1.8 $R_⊕$ planet, although this median does not take into account the decrease in eclipse depth for high impact factors.[5] If the BD is oblate and the planet's orbit is

---

[3] Also see Leconte et al. (2011). Beyond this article's title, this reference provides separate modeling of rotational deformation free from an exterior gravitational influence for BD mass range objects (J. Leconte 2013, private communication).
[4] Also see Eggl et al. 2012, *An Analytic Method to Determine Habitable Zones for S-Type Planetary Orbits in Binary Star Systems*
[5] Also note that Eq. 1 assumes the transit to start when the center of the planet touches the limb of the primary, whereas Eq. 2 assumes transit start at first contact. Assessing the error induced by all these imprecisions in modeling is beyond the scope of the present work.

**Table 1**

Photon noise-only signal-to-noise ratio (*S/N*) of the *detection* of a single transit event in *J* band for a non exhaustive list of BDs with $P_{\text{HZ out}} \leq 8$ h

|  | $P_{\text{HZ}}$ (h) in (Roche) | out | $\tau_{\text{min}}$ (min) | Depth (%) | mag (*J*) | photons s$^{-1}$ m$^{-2}$ (a) (× 10$^3$) | photons, per ½ transit @ HZ in (× 10$^3$) | *S/N* per transit | Geom. prob. HZ out (%) |
|---|---|---|---|---|---|---|---|---|---|
| SDSS J1416+13 B | 3.42 | 5.99 | 12.2 | 3.8 | 17.35[b] | 3.8 | 600 | 21 | 16 |
| 2MASS 0939-2448 A | 3.42 | 6.52 | 12.5 | 4.0 | 16.1[c] | 11 | 1800 | 34 | 14 |
| AB |  |  |  | 3.5 | 15.98 | 13.4 | 2200 |  |  |
| CFBDS J005910-011401 | 3.42 | 5.45 | 16.1 | 3.6 | 18.06[d] | 1.98 | 416 | 17 | 20 |

**Notes**

$P_{\text{HZ out}}$ is the orbital period at the outer edge of the radiative habitable zone. No tidal migration is considered here. 3.5 m-class telescope. We want the cadence to be half the minimal possible duration of the transit $\tau_{\text{min}}$ (at least one complete exposure taken during the transit).

[a] 0.82 transmission of the filter included. [b] Burningham et al. (2010). [c] The magnitude of the A component alone was estimated using values from Table 2, and it was checked that when doing the same with the B component the fluxes in the next column add up. [d] Scholz et al. (2009). For references on the remaining parameters of the targets, see Table 2.

aligned with the BD's spin, the duration of the eclipse will be larger.

Finally, the fast rotation of BDs separates the transit signal cadence from rotation-induced variability. BDs are thought to feature evolving inhomogeneities in their cloud deck (weather), likely to generate variability in a manner similar to that of star spots. The corotation distance is the orbit at which a planet's angular speed matches the BD's rotation. So as the rotation speed of the BD increases, the corotation distances moves inward with time. It becomes smaller than the Roche limit after 10-100 Myr, which is also the likely formation time-scale of terrestrial planets around a BD (Bolmont et al. 2011). Thus rotation-induced photometric variability of the BD primary would be in a totally different frequency regime than possible transit cadences. Still, if the *evolution time-scales* of the inhomogeneities are of the order of the transit cadence (planet orbital periods), rotation induced variability due to these features are very likely to increase the false alert rate (compare Fig. 3 with Roche orbital periods).

## 3. DETECTION STRATEGIES

### 3.1. The Importance of Completeness Assessment for Volume-limited Surveys

For any given primary, a photometric monitoring of a given sensitivity can yield a single self-significant transit event candidate[6] down to a planet radius $R_p$ (we consider a mean transit duration within our region of interest: the habitable zone). Monitoring this primary for a continuous duration $T$ implies that all potentially transiting planets with periods up to $T$, whatever their epoch (i.e. orbital phase), have been *screened* for. This can be dubbed '100% completeness up to $T$ (and down to $R_p$)'. Publishing the completeness of a volume-limited survey is particularly important when the final yield of detections is expected to be low, or even below unity. Indeed, the yield is necessary for planning fol-low-up eclipse spectroscopic characterization (including building of dedicated facilities).

---
[6] By default and unless mention otherwise no shift-adding of multiple observation nights (to increase the signal-to-noise ratio - *S/N* is considered in this paper (more on this in § 3.3).

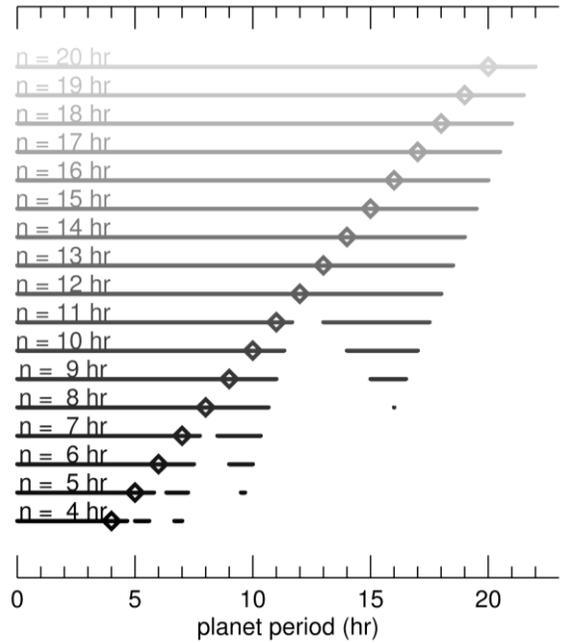

**Figure 4.** Planet orbital periods for which transit screening completeness is 100%, for two adjacent full monitoring nights each of usable duration $n$ (in the case of just one monitoring night the maximum orbital period for which completeness is 100% is of course only $n$ - diamonds). The step in $n$ between each grey shade is 1 hr, values for darkest and lightest are indicated.

We now examine specific strategies for different regions of the BD parameter space.

### 3.2. Ground based, Single Night, Single Longitude, 100% Completeness

If we ignore tidal migration (§ 4.2), the Roche-limited habitable zone (grey contour in Fig. 1) extends to orbital periods shorter than the typical photometric night durations (e.g. but not limited to, airmass < 2). The Roche limit is at 3.6 h for the 10 $M_\oplus$, 1.8 $R_\oplus$ super-Earth considered in Fig. 1, at 4.5 h for an analog of the Earth, and at 5 h for a 0.1 $M_\oplus$, 0.5 $R_\oplus$ planet. This means that some BDs can be screened with 100% completeness from the ground, in just a single observing night, with a single telescope at a single longitude – a remarkable efficiency.

When such a candidate is detected up to three subsequent follow-up nights are required: the first for determining a period and confirming the alert, a second for confirming the period and the periodic nature of the signal. An additional shifted third night can help ruling out submultiples of the initial period. A tradeoff in the shift has to be determined, since the greater the shift the greater the buildup of ephemeris uncertainties.

We explore the increase in orbital period screening by monitoring for one adjacent night (Fig. 4). For instance in the case of a 9 hr useful night the addition of one adjacent observation night yields 100% completeness for planet orbital periods up to 11 hr, and also for orbital periods between ~15 and ~16.5 hr (~39% increase).

Table 1 gives the expected, photon noise-only signal-to-noise ratio ($S/N$) of the *detection* of a single transit event with a 3.5 m-class telescope in the $J$ band for several nearby BDs with $P_{HZ\,out} \leq 8$ h. The integration time is derived from the minimal possible duration for the transit of a habitable planet (i.e. a transit at the inner edge of the habitable zone – HZ in). We further halve this integration time to take into account alternating between two bands (see the Artigau et al. observation mentioned above), in case no dichroic is available. We consider the overhead per one forth-back filter switching to be 40 s (case for *WIRCAM - Wide-field Infra-Red Camera* on the *Canada France Hawaii Telescope - CFHT*).

In the last column, we also give the geometric likelihood of transit at the *outer* limit of the habitable zone (i.e. the *lower* limit on the transit likelihood of habitable planets around these BDs; evidently, no primary types have higher habitable planet transit probability as the BD type).

To conclude, we note that BDs are currently being photometrically monitored with IR telescopes for increasingly extended continuous periods in the frame of atmospheric (weather) and evolution tracks research. We therefore call to this community to integrate the science case presented in this subsection in the evolution and further expansion of their field.

### 3.3 Ground, Multiple Nights

Brown dwarf habitable zones extend up to 10 days of orbital period (Fig. 2). BL08 suggested the use of a redundant, longitude distributed network of telescopes for continuous photometric monitoring, such as the *Las Cumbres Global Telescope* (*LCOGT*) network[7]. We have executed a test of such longitude distributed observation in early 2011. One $z = 17$ target was scheduled for 13.5 h of continuous monitoring, involving the two 2-m telescopes of the *LCOGT* in Hawaii and Australia and the 2-m *Himalaya Chandra Telescope* (*HCT*) with the *Himalaya Faint Object Spectrograph and Camera* (*HFOSC*).

Meteorological conditions enabled only observations from *HCT*, and the overall environmental conditions for that observation caused a high background level. Therefore the signal-to-noise ratio of the final light-curve (not shown) for this very faint target was too low for exploitation.

In conclusion, for the moment such a 2 m far red optical-class network may not be yet sufficiently longitude-redundant for robustness against environmental variability. Also, slightly brighter-on-average targets may relax the constraints on environmental conditions. However, these targets would have longer habitable zone outer limit periods, therefore requiring longer monitoring in order to achieve complete habitable zone screening. For instance already for the present test target the longest habitable period was longer than the 13.5 continuous hours we were able to secure. Longer monitoring means more different observatories are stringed together for such an observation. The red spectral energy distribution of BDs also advocates for extending the equipment of the 2 m class collectors worldwide with J-H-K detectors. Last, taking into account the meteorological forecasts at the different observatories and triggering the observing sequence in a Target-of-Opportunity (ToO) fashion could be investigated.

If the continuity is disrupted before the longest habitable period can be covered, a scheduling algorithm can enable to optimize the completeness of the screening of a given target (e.g. Saunders et al. 2009). The completeness may not reach near 100% but for a significant increase in observation time cost. However, such multiple observations enable to search for shallower transits and/or primaries with increased variability, by phase-folding search techniques, and accounting for the subsequent introduction of correlated (red) noise (von Braun et al. 2009). Note that flare-variability can be a real challenge for phase folding in M dwarfs light curves.

Last, Blake & Shaw (2011) have shown recently that, following the quality of the site, preciptable water vapor (PWV) variability can induce 5 mmag variations in $z$ band on the hour timescale; however they indicate that PWV can be monitored through Global Positioning System signals.

### 3.4. Intermediate Cases

For the BDs with outer habitable period between ~8 and ~20 h (i.e. the cases intermediate to those addressed in §§ 3.2 and 3.3 above), one would require a network such as the one described above, but operating in the infra-red. Such coordinated observations between telescopes usually operated through time allocation committees may prove difficult to set up (considering the very high pressure on these telescopes and constraints on mutual telescope observing coordination). Therefore, in the frame of a cohesive grand strategy for ground detection of habitable planets eclipsing BDs, coordinated observations are likely only as a second step, after 1 telescope-, 1 night observations on cooler targets are first demonstrated (§ 3.2 above). If no

---
[7] lcogt.net

**Table 2**
Total expected number of eclipsing habitable planets around nearby L, T and Y dwarfs.

| | Dist. (pc) | $T$ (K) | $M/M_\odot$ | $R/R_\odot$ | $L/L_\odot$ (log) | Ref. | $HZ_{Roche}$ | $HZ_{asymp.}$ 1 Myr | 10 Myr | trans. prob. 1 Myr | 10 Myr |
|---|---|---|---|---|---|---|---|---|---|---|---|
| WISE 1541-2250 | 2.8[a] | *350* | *0.011* | *0.01* | -6.88 | Cushing et al. 2011 | 0 | | | | |
| GJ 845 B a | 3.6 | *1320* | *0.065* | *0.0805* | *-4.699* | King et al. 2010 | 1 | | 0.64 | | 0.030 |
| b | | *910* | *0.050* | *0.0825* | *-5.232* | | 0.95 | | | | |
| SCR 1845-6357 B | 3.85 | *950* | *0.039* | *0.091* | -5.1 | Kasper et al. 2007 | 1 | | 0.27 | | 0.019 |
| UGPS 0722-05 | *4.1* | 505 | 0.005 | 0.10 | *-6.13* | Leggett et al. 2012 | 0.79 | | | | |
| DEN 0817-6155 | *4.9* | 950 | 0.04 | 0.089 | -3.53 | Artigau et al. 2010 | 0.98 | | | | |
| DEN 0255-4700 | 5.0 | 1300 | 0.035 | 0.09 | -4.62 | Stephens et al. 2009 | 1 | 0.55 | 0.74 | 0.023 | 0.033 |
| 2MASS 0939-2448 A | 5.3 | *700* | *0.038* | *0.085* | -5.8 | Leggett et al. 2009 | 0.52 | | | | |
| B | | *500* | *0.024* | *0.09* | -6.3 | | 0 | | | | |
| WISE 1741+2553 | 5.5[a] | | | | | [d] | 0.79 | | | | |
| 2MASS 0415-0935 | 5.7 | *947* | *0.01* | *0.12* | -5.0 | Del Burgo et al. 2009 | 1 | 0.24 | 0.14 | 0.021 | 0.011 |
| GJ 229 B | 5.8 | *950* | 0.038 | 0.094 | -5.2 | Geißler et al. 2008 | 1 | | 0.08 | | 0.006 |
| GJ 570 D | 5.9 | *948* | 0.019 | 0.11 | -5.0 | Del Burgo et al. 2009 | 1 | 0.55 | 0.14 | 0.051 | 0.010 |
| SIMP J013656.5+093347.3 | 6[b] | *1200* | 0.044 | 0.097 | -5.25 | Artigau et al. 2009 | 0.96 | | | | |
| 2MASS 0937+2931 | 6.1 | *950* | *0.054* | 0.08 | -5.33 | Leggett et al. 2010 | 0.87 | | | | |
| WISE 0254+0223 | 6.1[a] | 660 | 0.01 | 0.11 | -5.7 | Kirkpatrick et al. 2011 | 0.95 | | | | |
| WISE 1738+2732 | 7[c] | 350 | *0.019* | *0.093* | -6.94 | Cushing et al. 2011 | 0 | | | | |
| | | | | | | | | | | 0.09 | 0.11 |
| | | | | | | | | | **Expected #** | | |
| | | | | $\times 0.41\,^{+0.54}/_{-0.13}$ = | | **0.78** $^{+1}/_{-0.25}$ | | | **0.036** $^{+0.048}/_{-0.012}$ | **0.078** $^{+0.06}/_{-0.014}$ |
| | | | | | | | | | **% probability of at least 1 occurrence** | | |
| | | | | | | | | | **56** $^{+31}/_{-13}$ | **3.9** $^{+4.9}/_{-1.2}$ | **4.5** $^{+5.6}/_{-1.4}$ |

**Notes**

$HZ_{Roche}$ is the fraction of circular orbits that are in the habitable zone but outside a 10 $M_\oplus$ Roche limit (uniform distribution in radius). Similarly $HZ_{asymp}$ is the fraction, from the remaining habitable zone, where planets can exist at the end of the tidal migration process, for two different planet formation ages. The last two columns give the corresponding transit probability of the median orbit in the remaining final effective HZ, weighted by $HZ_{asymp}$. Therefore, the total of these two last columns (0.09 and 0.11), multiplied by $\eta_\oplus = 0.41\,^{+0.54}/_{-0.13}$ from Bonfils et al., are estimates of the total expected number of habitable planets transiting BDs in this volume (see text for detailed justification).

Distances (**pc**) are RECONS parallaxes unless mentioned otherwise. When parallaxes were not available, we use photometric distances. Values in italic are from the reference. The remaining non-italic parameters (among photospheric temperature $T$, mass $M$, radius $R$ and luminosity $L$) are interpolated from COND03 grids using the parameters from the reference. Note that the purpose of this table is to compute some ensemble averages; therefore the values of the BD parameters should not be reused for the study of individual objects, since the uncertainties on most parameters are quite large (e.g., spectroscopic/photometric distance estimates), and because of ongoing refined observations (e.g. parallaxes).

[a] Kirkpatrick et al. 2011, parallax. [b] Faherty et al. 2009. [c] Average of quite dissimilar photometric distance (Kirkpatrick et al. 2011) and spectroscopic distance (Cushing et al. 2011). [d] For this BD, interpolation of the grids to the 2MASS $J$ and $H$ magnitudes (Kirkpatrick et al. 2011, Scholz et al 2011) did not converge (as it was more the case for WISE J0254+0223). We therefore use here the values of the only other T9 BD in the sample, UGPS 0722-05.

coordinated observation can be set up observations have to be spread throughout the observing season of the target, arranging them so that together they satisfactorily cover the time-folded range of orbits that is sought, significantly increasing the total cost in telescope time. See also Berta et al. (2012) for a related study deriving from the MEARTH survey for habitable planets transiting M dwarfs. This study includes analytical tools for integrating "lone transit events" (from different telescopes using different filters at different observatories) into coherent planet candidates.

The optimal approach for screening these intermediate cases is a dedicated monitoring program from space, where uninterrupted monitoring can be achieved. Since 2011 August the Spitzer Warm Mission Exploration Science Program 80179 "*Weather on Other Worlds: A Survey of Cloud-Induced Variability in Brown Dwarfs*" (Metchev: PI) is monitoring one after another, for a minimum of 21 continuous hours each, BDs from a list of 44 targets (873 hr awarded in total)[8]. Unfortunately none of the 25 targets observed until now are at or beyond than 7 pc. Also observing simultaneously in both channels of Warm Spitzer (3.6 and 4.5 μm) is not possible because the arrays of each channel see different non-overlapping parts of the sky. Therefore a prospective interlaced mode is not up for con-sideration at the cadence required both by intrinsic variability studies or exoplanet detection.

---

[8] http://sha.ipac.caltech.edu/applications/Spitzer/SHA//#id=SearchByProgram&DoSearch=true&SearchByProgram.field.program=80179

## 4. EXPECTED NUMBER OF NEARBY ECLIPSING HABITABLE PLANETS

### *4.1. Previous Study*

Due to *S/N* constraints, habitable planets can be searched for biosignatures with eclipse spectroscopy only out to a limited distance. Belu et al. (2011, Fig. 8) find that at 6.5 pc emission biosignatures can be detected within *JWST*'s baseline lifetime only around primaries of spectral types later than ~M5. The *S/N* scales as the planetary radius squared, and inversely with the distance. Also, considering a specific spectral signature a given planet might exhibit (and which would require to be confirmed or rejected), the signature's *S/N* scales linearly with the strength of the spectral feature (number of atmospheric scale heights for primary eclipse and brightness temperature depth for secondary), and with the inverse of the resolution.

Belu et al. further showed (Fig. 19 therein) that in a ~7 pc volume[9], the total expected number of habitable eclipsing planets is ~0.3 for the M5-M9 dwarf primaries population (considering the recent Bonfils et al. (2011) lower bound for $\eta_\oplus$: 0.41). Note that Selsis et al. (2007a) pointed out that a primary may host in principle several habitable planets (as was perhaps the case for the Sun 4 Gyr ago, when Venus, the Earth and Mars were potentially habitable). Therefore the case of a final $\eta_\oplus > 1$ is not to be discarded. For the compact habitable zones of BDs ($10^{-3}$ AU scale), further studies are needed to address the stability of multiple planets in this zone, and their possibility to form and/or migrate (see next subsection).

### *4.2. Present Study*

We now extend the previous study to T, L and Y dwarfs; some young BDs could exhibit M spectral type (and vice-versa): they would have been included in the previous study. Table 2 lists the known BDs likely within 7 pc. This table was compiled from RECONS (recons.org), SIMBAD and other recent discoveries. For each object we have collected from the literature estimates of photospheric temperature, mass, radius and luminosity. The missing values were interpolated from the COND03 evolutionary grids (Baraffe et al. 2003) using the available parameters.

Table 2 contains two binary systems. GJ 845 Bab (ε Indi Bab) is a binary BD system, with a separation between the components of at least 2.1 AU (Volk et al. 2003). The outer limit of the habitable zone for each of the two components (S-type orbits) is 0.01 and 0.005 AU respectively, so orbital stability at these distances is not at issue. On the other hand 2MASS 0939-2448 is likely a dissymmetric close binary (Burgasser et al. 2008, Leggett et al. 2009) with a separation of under 0.03 AU. Holman & Wiegert (1999) have derived semi-major axis upper limits for dynamically stable orbits around each component of a binary. The outer habitable orbits around each likely component of 2MASS 0939-2448 (0.0029 and 0.0016 AU respectively - or 7.2 and 3.8 times the radii of their respective primaries) are indeed dynamically stable.

---

[9] Volume limit of the 100 nearest objects at that time, RECONS (www.recons.org)

We then calculate for each BD the fraction of circular orbits that are in the habitable zone but outside the Roche limit (column $HZ_{\text{Roche}}$, 10 $M_\oplus$ planet).

When computing geometric transit probabilities it is usually assumed that the probability of a planet occupying a given orbit is flat across the available parameter space, e.g., the HZ. However, given the close proximity of BDs' HZs, tidal interactions between the planet and BD act to modify a planet's orbit after its formation, with consequences for its transit probability.

Bolmont et al. (2011) studied the tidal evolution of planets orbiting BDs (previously referred to here as *tidal migration*). The basic concept is as follows. We already mentioned the gradual reduction of the corotation distance (§ 2). This is important because a planet's position with respect to corotation determines the direction of tidal migration. Given that BDs' corotation distances shrink in time, almost all planets that survive around BDs experience outward tidal migration. Thus, there is a parameter-dependent orbital radius inside of which planets should not exist. The asymptotic limit is reached after 10-100 Myr. The ages of observed BDs are known with precisions equal to or larger than 0.1 Gyr, and all of the BDs in Table 3 have age estimates larger or equal to 100 Myr. We therefore proceed and apply the asymptotic limit model to all the BDs in Table 2.

The key parameters that determine this limit are the BD and planet masses and their internal dissipation rates (Bolmont et al. 2011). From this reference we use Figs. 6 and 8 which give the asymptotic limit (semi-major axis) for planets around BDs of different masses (for a 1 $M_\oplus$ planet forming at 1 and 10 Myr respectively). And we give for the remaining BDs in Table 2 that have $HZ_{\text{Roche}} > 0$ (3 BDs do not) the non null fractions of circular orbits that are in the habitable zone but with semi-major axis larger than the asymptote (column $HZ_{\text{asymp.}}$).

Finally, for the remaining BDs that still feature the above effective final habitable zone, we compute the transit likelihood at the middle of the zone (with the same working assumption $\eta_\oplus = 1$). And we weight (multiply) this likelihood by $HZ_{\text{asymp.}}$ thus obtaining a final transit probability (column *transit. prob.*). The justification for this weighting is the following. These individual likelihoods are then multiplied by the recently directly measured $\eta_\oplus$ within the M dwarf primary population, $0.41^{+0.54}_{-0.13}$ (Bonfils et al. 2011, lower limit, as per sensitivity of the technique to the whole range of habitable planet masses). Weighting our individual transit likelihoods by $HZ_{\text{asymp.}}$ is a valid approach because the habitability of individual planets in the sample of Bonfils et al. is based on radiative (from the primary) con-siderations alone, and their primary population is not subject to the limitations included in the $HZ_{asymp.}$ factor (Roche limit and tidal-migration). Their $\eta_\oplus$ contains only information on planet orbital density as a function of the primary; it can validate formation and migration models, the latter excluding tidal migration because the habitable distances around M dwarfs are too large for this mechanism. It is this formation- and migration-other-than-tidal- planet density function of Bonfils et al. that we extrapolate here to BDs. We must caution however that the figures in Bolmont at al. (2011) seem to indicate that the tidal migration mechanism tends to redistribute orbits (i.e. change the density of orbits, either shepherding together or dispersing, depending on the initial conditions).

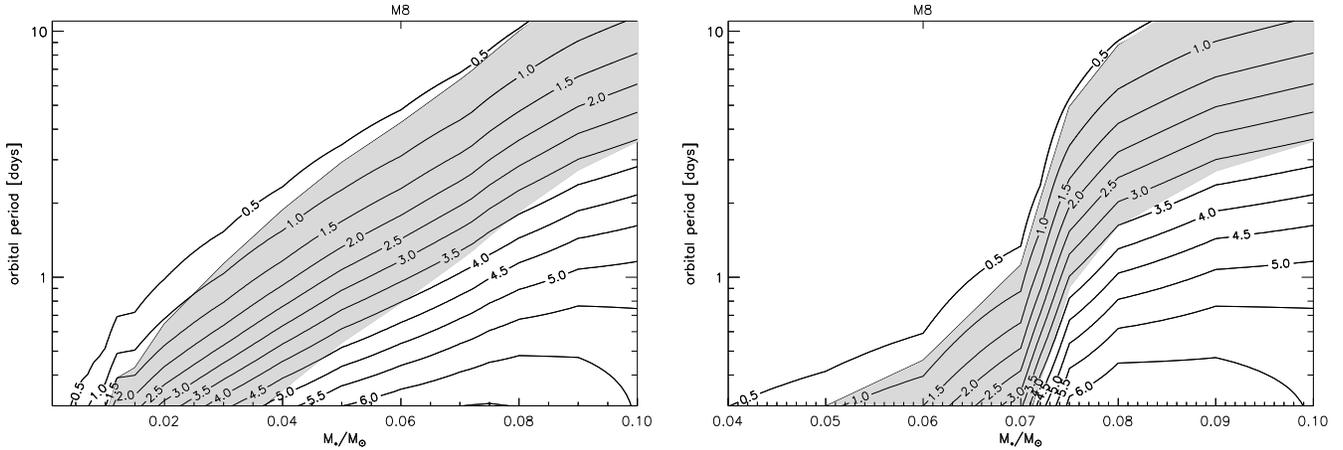

**Figure 5.** Signal-to-noise ratio (*S/N*) on the detection of a spectral feature in emission (secondary eclipse spectroscopy), at 10 μm, as function of the mass of the brown dwarf and the orbital period of the planet. The spectral feature has a brightness temperature depth of 30 K and is 0.1μm wide (i.e. $R = 100$). The planet is a 1.8 Earth-radii super-Earth, and the system is situated at 6.7 pc. The observations of 90 eclipses are summed for this result. The age of the brown dwarf is 1 Gyr (**left**) and 10 Gyr (**right, note the different abscissa scale**). The **grayed** area is the habitable zone (§ 2). The *S/N* scales linearly with the square root of the number of summed eclipses (if no correlated noise), with the square of the planet's radius, with the brightness temperature depth, with the inverse of the distance in pc, and with the inverse of the resolution.

The final expected number of eclipsing habitable planets around BDs within 7 pc is given in bold at the bottom of Table 2. A more significant number is the probability for a survey to yield at least one transiting habitable planet in this volume (also in Table 2). Assuming the optimistic scenario of late planet formation around BDs (i.e. at 10 Myr), the survey of the corresponding 6 BDs in Table 2 has $4.5^{+5.6}/_{-1.4}$ % chance of yielding at least one habitable eclipsing planet.

Given the recent nature of the work on tidal migration, we also include the significantly more optimistic figures when tidal-migration is not considered. The probability for the survey of the above 14 BDs with Roche limited-only HZs to yield at least one transiting habitable planet likely within 7 pc is then $56^{+31}/_{-13}$ %.

*4.3. Discussion*

Thus, to include BD primaries in the search for nearby, eclipse-characterizable habitable planets is to increase the expected number of occurrences ~2.5-fold (when compared with the late M dwarf-only search).

The results on tidal migration depend on some parameters which are unknown/poorly constrained for our nearby BDs, such as the dissipation factor in the BD or the initial rotation rate. There are also various rotation braking mechanisms that are not considered. This is to be combined with the uncertainties on the parameters of which we give the estimates (mass, age). Will future refinements of tidal migration modeling enable to gain back the order of magnitude between the Roche-only limitation of the habitable zone and the one by tidal migrations? Or will it completely rule out habitable planets around BDs? What about other mechanisms for late migration, or about the frequency of late planet scattering?

The planet's mass is also a strong factor for tidal migration, with planets of 0.1 $M_\oplus$ (0.5 $R_\oplus$) hardly experiencing any effect. Unfortunately the performance of the subsequent search for spectral signatures scales with the square of the planet's radius. But this should also remind us that the volume limit we used for our list is an average estimation of spectroscopic characterization capability with the *JWST*; particularly favorable cases (strong spectral signatures, planets close to the inner limit of the habitable zone, etc.) may be characterizable further away. Planet detection surveys should therefore plan a significant margin on these volume estimates (and the number of targets scales with the cube of the distance).

Also note that only lower bounds on the local space density of BDs are presently available (Kirkpatrick et al. 2011). The final BD detection count from the ongoing processing of the *Wide-field Infrared Survey Explorer* (*WISE*) data is expected to be ~1,000 BDs, which should double or triple[10] the number of know primaries within 25 light years (7.6 pc). Our values should be considered therefore as lower limits.

---

[10] As of December 2009, NASA *WISE* Launch Press Kit.

## 5. FUTURE CHARACTERIZATION

We now consider habitable planet secondary eclipse spectroscopy performance around a BD. We reprise our previous such study around F-M dwarfs with the *JWST* (Belu et al. 2011 for detailed description of the modeling). Brown dwarfs may not exhibit significant near-UV flux. Therefore a (biotic) $O_2$ atmosphere on a BD exoplanet may not generate $O_3$ (ozone) in its stratosphere, which is a convenient $O_2$ detection proxy around 10 μm (location of thermal emission from a body at habitable temperatures). The question of BD HZs and biosignatures is discussed at length in § 6.2 & 6.3. We therefore consider a fiducial spectral feature in emission (at 10 μm, brightness temperature depth of 30 K, and 0.1 μm wide - i.e. resolution $R = 100$).

The instrument considered is the *Mid Infra-Red Instrument* (*MIRI*) in Low Resolution Spectroscopy (LRS) mode. For 1 and 10 Gyr-old BDs, Figure 5 shows the signal-to-nose ratio (*S/N*) on the *detection* of our fiducial spectral feature from a 1.8 Earth-radii planet at 6.7 pc, summing the observations of 90 secondary eclipses. Program time cost per eclipse is twice eclipse duration (at least ~30 min, Fig. 2), plus the 65 min generic *JWST* slew time budget, every 10-70 h (period of the planet). Note that even for the longest period planets (~10 days), 90 transits are well within the telescope's lifetime. We reprise here our comment from Belu et. al (2011): such an hypothetical observation, which would happen on only one (see § 4) most interesting transiting system, represents a total telescope time only a magnitude larger than the longest exposures made until now with the Hubble *Space Telescope* (*HST,* Beckwith et al. 2006). Also note that the 1.8 Earth radii is an upper limit for habitability, but could be extremely optimistic in terms of initial mass available in a BDs protoplanetary disk for planet formation.

Despite the lower luminosity of the primary, hence the reduced photon noise, shorter orbital periods also mean shorter occultation durations (10-40 minutes). This curbs the gain one could have expected relative to the case around M dwarfs. The discussion at the end of § 2 on the rotation-induced variability of primary and transit detection also applies this eclipse characterization follow-up.

One can see that atmospheric absorption features such as the one presented here can be detected on habitable planets eclipsing BDs after a follow-up of a couple of months, for a cost < 2.5 h of observation every $1 - 2$ days, so on average about 1% of the 5-year mission time of the *JWST*. This cost in mission time is about a factor 2 better on average than for M dwarf habitable planets (Belu et al. 2011), and more importantly, spread over only 1/5th to 1/10th of the 5-year mission time (whereas in the M dwarf case the required number of observations spreads over the entire mission life-time, exposing to the risk of dedicating time and acquiring data that ends up having insufficient *S/N* if the JWST were to become inoperable too soon).

## 6. DISCUSSION

### 6.1. Terrestrial Planet Formation and Orbital Evolution around BDs

BL08 and Bolmont et al. (2011) reviewed the literature on the likelihood of formation of habitable planets around BDs. There is ample evidence in favor of terrestrial planet formation around BDs: the same fraction of young BDs has circumstellar disks as do T Tauri stars (Jayawardhana et al. 2003, Luhman 2005), and there is observed evidence of grain growth in BD disks (Apai et al. 2005). Of course, the exact outcome of the accretion process depends on the disk mass and mass distribution (Raymond et al. 2007, Payne & Lodato 2007), which probably scales roughly linearly with the primary mass (Andrews et al. 2010). Regarding formation, see also Charnoz et al. (2010) for late accretion at the Roche edge of a debris disk.

If tidal migration influences are confirmed there should be no planets with orbital period under 8 h orbiting them (Fig 3, black contours). Only extreme unlikely scenarios could allow such planets, like unusually low dissipation factors, unusually high initial rotation rate, or very recent capture, migration or formation. For instance Fig. 19 in Bolmont et al. (topmost panel for the lowest BD dissipation factor) shows a tremendous sensitivity to the initial semi-major axis when computing the final asymptotic one. In theory there could be an extremely narrow interval of initial positions for which the final asymptotic orbit is as close as desired to the Roche limit. This interval is likely $\ll 10^{-3}$ AU (the span of initial semi-major axes for which planets that migrate in are saved, whatever their final asymptotic semi-major axis).

However, we recall that the detection of hot jupiters was proposed as feasible almost half a century before their detection (Struve 1952). To summarize, there are significant theoretical uncertainties associated with each of these questions and no certain answers at the current time. We are of the opinion that, given their exceptional detection advantages (1 night-, 1 location- 100% completeness) we should invest in the presented strategy. Additionally, whole night monitoring of BDs may enable improved BD atmospheric studies. Therefore planets transiting BDs with periods under 8 h have a high *payoff / screening cost* ratio.

### 6.2. Habitability and BDs

A radiative habitable zone (HZ), within which terrestrial planets can sustain surface liquid water, can be defined around BDs. The inner-edge of the HZ corresponds to a $H_2O$-rich atmosphere and the outer edge to a greenhouse efficient gas-rich atmosphere – most likely $CO_2$. The inner limit is reached when the mean stellar flux absorbed by the planet is 300 W m$^{-2}$ (runaway greenhouse threshold). Determining the location of the outer edge, which depends on the efficiency of $CO_2$ as a greenhouse gas, will require specific climate modeling (1D & 3D), due to the strong overlap between the thermal emission of the BD and the molecular lines in the planet's atmosphere (e.g., Words-worth et al. 2011). The full absorption of the continuum of $H_2O$ and $CO_2$, and the absence of Rayleigh scattering will likely lead to a planetary albedo close to null (we take 0.1 in § 5). Strong stratospheric warming and inefficient greenhouse is expected, possibly leaving the surface at a lower temperature than the globally perceived brightness temperature of the stratosphere. We are currently developing 1D and 3D codes suitable for BD planets. The contribution of internal heating due to tidal effects (Jackson et al. 2008 for stellar masses down to 0.1 $M_\odot$) may also be significant.

Several threats against surface habitability exist within the HZ of BDs. One is the tidal spin-orbit synchronization. Planets on circular orbits inside the HZ of M stars and BDs are expected to have a permanently dark hemisphere and a zero obliquity. If zonal and meridional heat transport is insufficient the water and the atmosphere can end as condensed caps at the poles and night side of the planet. Simulations done for GJ 581 d (Wordsworth et al. 2011) show however that a dense atmosphere can provide enough heat transport to homogenize the temperature over the whole surface (see also Joshi et al. 2003).

Another threat is the rapid cooling of the BD (Fig. 5), which has two implications. The first one is that, coupled with the tidal migration outward drift, a planet remains habitable during only a fraction of the BD's life (more than 1 Gyr only for BDs > 0.04 $M_\odot$ - Bolmont et al. 2011). Nonetheless, life on Earth is thought to have existed within 1 Gyr of its formation. Thus, although planets have a short habitable window around BDs, it is of great scientific interest to search for them (see Lopez et al. 2005 for a similar discussion about the HZ around red giants). The second implication is that habitable planets were initially on the hot side of the HZ. Around a Sun-like star such a hot location would imply atmospheric losses of water. Venus for instance has kept little of its initial water reservoir, as shown by the high D/H ratio of its remaining water (a few tens of cm precipitable). If planets lose most of their water content during their pre-habitable history, they are unlikely to become habitable worlds when the HZ catches up with them. The case of Venus, however, may not be a relevant analog for BD planets. Indeed, the Sun emits significant UV and XUV(EUV) fluxes, respectively able to photolyse $H_2O$ and to drive the atmospheric escape by heating the exosphere.

It is unclear whether BDs have enough activity to produce such fluxes that would result in a significant water loss. For G, K and early M stars, magnetic activity and resulting XUV emission is correlated with rotation rate (Ribas et al. 2005; Scalo et al. 2007) and thus the XUV levels in the HZ of early M stars remain very high (higher than in the HZ of the Sun) for 1 to a few Gyr. For late M stars and BDs, and despite their high rotation rate, there is a steep drop-off of activity, which may be explained by their lower atmospheric temperature and ionization fraction (Mohanty et al. 2002). Very young BDs do exhibit observable X rays (Preibisch et al. 2005) but likely to come from the accretion of a protoplanetary disk ergo predating the formation of planets. Therefore in the absence of signifi-cant photolysis and exospheric heating, it is possible for planets on the hot side of the HZ of a BD to keep a steam atmosphere long enough to become habitable. Note also that, even in the case of significant atmospheric and water erosion, the amount of water that remains for the habitability window depends on the initial reservoir. Volatile-rich planets, or so called ocean-planets, that have formed in the cold outer part of the protoplanetary disk and migrated toward inner regions, can keep more than a terrestrial ocean for Gyrs even if located close to a Sun-like star (Selsis et al. 2007b). It is therefore possible to have oceans at the surface of planets in the HZ of BDs.

**Note**: Since the submission of this article, Barnes & Heller (2013) have independently addressed & quantitatively furthered some of the questions raised above in this Section; especially they include the potential effect of tidal heating when the planet eccentricity is forced to non-zero values by planet-planet interactions.

### 6.3. Biosignatures from BD planets

The atmospheric biosignatures paradigm rests on the ability to detect an out-of-equilibrium thermodynamical state and that and that all simpler physical processes fail to reproduce this state. For instance the photolysis and escape mentioned above can lead to abiotic $O_2$ buildup. Unfortunately a back of the envelope calculation shows that a detectable level of UV emission from nearby BDs (with the *HST*) corresponds to levels in their habitable zone significantly higher that the ones required for such $O_2$ buildup.

Supposing that all simpler physical causes for the observed out-of-equilibrium state are ruled out it is then likely that a more complex process is responsible. On Earth it is oxygenic photosynthesis (and not chemoautotrophy) that is responsible for the out-of-equilibrium state of our atmosphere (Rosing 2005). Oxygenic photosynthesis uses photons to break water molecules and the liberated hydrogen to reduce $CO_2$. The water bonding energy corresponds to a wavelength of 240 nm. As no such short wavelength reaches the surface of the Earth life has adapted to the available, longer wavelengths spectrum and eventually managed to store the energy of the number of photons required for breaking a single $H_2O$ molecule. Using the even longer wavelengths of a BD may imply only to store more photons per $H_2O$ molecule. Actually, photosynthesis may have evolved from initial infrared sensors used to detect sources of heat (Nisbet et al. 1995), and there are claims that photosynthetic organisms still use infrared light at great oceanic depth (Beatty et al. 2005). There is, therefore, no reason to rule out the possibility of a photosynthetic activity (oxygenic or not) evolving on a BD planet.

So going back to BDs possibly lacking photochemical out-of-equilibrium influence on their HZ planets, detection of biosignatures would be a most robust detection of life.

Remains the question of how to detect $O_2$? BDs do not provide enough flux at 760 nm for primary eclipse spectroscopy detection (except perhaps for the brightest BDs). Also no $O_3$ buildup is expected (§ 5). Therefore detecting $O_2$ in the absence of significant photochemistry and visible flux seems to be challenging. One path could be to investigate the observability of the dimers $O_2$-$X_2$ ($O_2$-$O_2$, $O_2$-$N_2$, etc.) at 1.26 µm, in the extended column at the limb during primary transits (Pallé et al. 2009 – Fig. S3).

## 7. CONCLUSION

In this paper we have examined three different strategies for detecting eclipsing habitable planets around BDs, depending on the maximum habitable orbital period. These group into ground searches and into space based searches (*Spitzer Space Telescope*).

Planets orbiting massive and old BDs can have their maximum habitable orbital period shorter than the duration of the observing night. Though they may be rare they can be screened for with 100% completeness in only one (1) night, from a single location (1 telescope). Conducting the transit search in the near-infrared is mandatory.

We have started investigating the monitoring of bright BDs with 2m-class telescope network (multiple institutions) in the deep red optical. Coordinated observation with a longitude-distributed network of telescopes can increase the duration of the continuous monitoring sequences therefore reducing the time-cost for achieving satisfactory screening completeness for a given target. However weather statistics at the various sites (and their correlation) should be included in estimating the time-cost of such campaigns, and the need for additional longitude-redundant coordination. It is the third option, of monitoring from space, which appears as the most effective.

We also show that the density of habitable eclipsing planets around BDs vary greatly depending on mechanisms that have only recently started to be investigated around BDs. Consider a survey of the habitable orbits for the 21 closest BDs (7 pc). The likelihood to detect at least one (habitable) transiting planet varies between $4.5^{+5.6}_{-1.4}$ and $56^{+31}_{-13}$ %, depending on whether tidal evolution is taken into account or not. Even in the pessimistic case, since these planets are also remarkably easy to find if do exist, a search program is worth the risk since it can validate these newly investigated mechanisms (tidal-induced migration).

Occultation spectroscopic characterization of a habitable planet around a BD within 5 to 10 pc is achievable with the *JWST*. Such a program would be spread over only 1/5$^{th}$ to 1/10$^{th}$ of the mission's life-time (instead of the whole mission for planets around M dwarfs).

More generally, given uncertainties in the existence of available nearby eclipse-characterizable habitable planets, an effort should be made to upgrade transit surveys around low mass dwarfs projects to increased sensitivity (as planed in recent Berta et al. (2012), but also increased collector diameter, better filters, better site). Setting up new surveys is also to be considered. The completeness of the screening for eclipsing habitable planets around nearby low mass dwarfs should be published at the earliest, to enable if required the adjustment of roadmaps toward their characterization in combined light. In case of absence of eclipsing planets, the effort toward characterization should once again be fully redirected to the spatially resolving track mentioned in the introduction.[11] The interest of a local sample is not only in terms of astronomical photon $S/N$; but also inspirational, for future generations (e.g., probe sending, see also Belu 2011).

We are most thankful to Amaury Triaud, Michael Gillon, Roi Alonso & Monika Lendl for helping with this work. AB acknowledges support from CNES. FS acknowledges support from the European Research Council (ERC) starting grant 209622: $E_3ARTH^s$. SNR thanks the CNRS's PNP program. PF acknowledges support from the ERC/European Community under the FP7 through Starting Grant agreement number 239953, as well as by Fundação para a Ciência e a Tecnologia (FCT) in the form of grant reference PTDC/CTE-AST/098528/2008. This research made use of www.solstation.com, Aladin and JSkyCalc. AB thanks J.-F. Lecampion for assistance in light curve production. We thank Ludovic Puig for contributing to the inspiration for this work. We thank the anonymous referee for attentive reading that helped improve the manuscript, and the numerous educational and prospective comments.

## REFERENCES

Abbot & Switzer 2011, ApJ, 735L, 27
Andrews, S. M., Wilner, D. J.; Hughes, A. M., Qi, C., Dullemond, C. P. 2010, ApJ, 723, 1241
Apai, D., Pascucci, I., Bouwman, J., et al. 2005, Science, 310, 834
Artigau, E., Bouchard, S., Doyon, R., & Lafrenière, D. 2009, ApJ, 701, 1534
Artigau, E., Radigan, J., Folkes, S., et al. 2010, ApJ, 718L, 38
Baraffe, I., Chabrier, G., Barman, T. S., Allard, F., & Hauschildt, P. H. 2003, A&A, 402, 701
Beatty, J. T., Overmann, J., Lince, M. T., et al. 2005, PNAS, 102, 9306
Beaulieu, P., Kipping, D. M.; Batista, V. 2010, MNRAS, 409, 963
Beckwith, S. V. W. 2008, ApJ, 684, 1404
Barnes, R. & Heller, R. 2013, Astrobiology, accepted
Belu, A. R., Selsis, F., Morales, J.-C. et al. 2011, A&A, 525A, 83
Belu, A. R. 2011, in Springer, Studies in Space Policy, 5, ESPI & ESF – Landfester, U,, Remuss, N.-L., Schrogl, K.-U., & Worms, J.-C. (eds.), 57-64
Berta, Z., Irwin, J., Charbonneau, D., Burke, C., & Falco, E. E. 2012, AJ, 144, 145
Blake, C. H., & Shaw, M. M., 2011, PASP, 123. 1302
Blake, C., Bloom, J. S., Latham, D. W., et al. 2008 PASP, 120 860
Bolmont, E., Raymond, S. N., & Leconte, J. 2011, A&A, 535A, 94
Bonfils, X., Delfosse, X., Udry, S., et al. 2011 arXiv:1111.5019
Burningham, B., Leggett, S. K., Lucas, P. W. et al. 2010, MNRAS, 404, 1952
Burgasser, A. J.; Tinney, C. G., Cushing, M. C., et al. 2008 ApJ 689L 53
Caballero, J. A. & Rebolo, R. 2002, F.Favata, I.W. Roxburgh & D. Galadí Enríquez eds., ESA SP-485
Caballero, J. A. 2010, Highlights of Spanish Astrophysics V, 79
Charbonneau, D., Irwin, J.; Nutzman, P., & Falco, E. E. 2008, AAS, 40, 242

---

[11] Combined light spectroscopic characterization of non-eclipsing habitable super-Earths may be achievable with the *JWST* for extremely favorable (e.g., amongst other, most nearby) cases - see preliminary work by Selsis et al. (2011). Completeness of the screening for this population may be achieved by a mission like the NEAT proposal (Malbet et al. 2011)

Charnoz, S., Salmon, J., Crida, A. 2010, Nature, 465, 752
Cockell, C. S., et al., 2009, Astrobiology, 9, 1
Cushing, M. C., Kirkpatrick, J. D., Gelino, C. R., et al. 2011, ApJ, 743, 50
Del Burgo, C., Martín, E. L., Zapatero Osorio, M. R., & Hauschildt, P. H. 2009, A&A, 501, 1059
Deming, D. L., Seager, S., Winn, J., et al. 2009, PASP, 121, 952
Eggl, S., Pilat-Lohinger, E., Georgakarakos, N., Gyergyovits, M., & Funk, B. 2012, ApJ, 752, 74
Faber, J. A., Rasio, F. A., & Willems, B. 2005, Icarus, 175, 248
Faherty, J. K., Burgasser, A. J.. Cruz, K. L. et al. 2009, AJ, 137, 1
Ford, E. B., & Rasio, F. A. 2006, ApJ, 638, L45
Geißler, K., Chauvin, G., & Sterzik, M. F. 2008, A&A, 480, 193
Grillmair, C. J., Burrows, A., Charbonneau, D., et al. 2008, Nature, 456, 767
Herbst, W., Eislöffel, J., Mundt, R., & Scholz, A. 2007, Protostars & Planets V, 297
Holman, M. J. & Wiegert P. A. 1999, AJ, 117, 621
Jackson, B., Barnes, R., & Greenberg, R. 2008, MNRAS, 391, 237
Jayawardhana, R., Mohanty, S., & Basri, G. 2003, ApJ, 592, 282
Joshi, M. 2003, Astrobiology 3-2, 415
Kaltenegger, L., & Traub, W. A. 2009, ApJ, 698, 519
Kasper, M., Biller, B. A.; Burrows, A., et al. 2007, A&A, 471, 655
King, R. R., McCaughrean, M. J., Homeier, D., et al. 2010, A&A, 510A, 99
Kirkpatrick, J., Cushing, M. C., Gelino, C. R., et al. 2011, ApJSS, 197, 19
Knapp, G. R., Leggett, S. K., Fan, X., et al. 2004, AJ, 127, 3553
Lammer, H., Selsis, F., Chassefière, E. et al. 2010, Astrobiology 10-1, 45
Leconte, J., Lai, D., & Chabrier, G. 2011, A&A, 528A, 41
Leggett, S. K., Burningham, B., Saumon, D. et al. 2010, ApJ, 710, 1627
Leggett, S. K., Cushing, M. C.; Saumon, D., et al. 2009, ApJ, 695, 1517
Leggett, S. K., Saumon, D., Marley, M. S., et al. 2012, ApJ, 748, 74
Lopez, B., Schneider, J., & Danchi, W. C. 2005, ApJ, 627, 974
Luhman, K. L., Adame, L., D'Alessio, P., et al. 2005, ApJ, 635, L93
Malbet, F., Léger, A., Shao, M., et al. 2011, ExA Online First, 109
Mohanty, S., Basri, G., Shu, F., Allard, F., & Chabrier, G. 2002, ApJ, 571, 469
Nisbet, E. G., Cann, J. R., & Van Dover, C. L. 1995, Nature, 373, 479
Paczyński, B. 1971, ARA&A, 9, 183
Pallé, E., Zapatero Osorio, M. R., & García Muñoz, A. 2011, ApJ, 728, 19
Pallé, E.; Zapatero Osorio, M. R., Barrena, R. 2009, Nature, 459, 814
Payne, M. J., & Lodato, G. 2007, MNRAS, 381, 1597
Preibisch, T., McCaughrean, M. J.; Grosso, N., et al. 2005, ApJS, 160, 582
Rauer, H., Gebauer, S.. Paris, P. V., et al. 2011, A&A, 529A, 8
Raymond, S. N., Scalo, J., & Meadows, V. S. 2007, ApJ, 669, 606
Ribas, I., Guinan, E. F., Güudel, M., & Audard, M. 2005, ApJ, 622, 680
Rosing, M. T. 2005, International Journal of Astrobiology, 4, 9
Saunders, E. S., Naylor, T., & Allan, A. 2008, AN, 329, 321
Scalo J., Kaltenegger, L., Segura, A., et al. 2007, Astrobiology, 7-1, 85
Selsis, F., Chazelas, B., Bordé, P., et al. 2007b, Icarus, 191, 453
Selsis, F., Kasting, J. F., Levrard, B., et al. 2007a, A&A, 476, 1373
Selsis, F., Wordsworth, R. D., & Forget, F. 2011, A&A, 532A, 1
Scholz R.-D., Bihain, G., Schnurr, O., & Storm, J. 2011, A&A, 532L, 5
Scholz R.-D., Storm, J., Knapp, G. R., & Zinnecker, H. 2009, A&A, 494, 949
Stephens, D. C., Leggett, S. K., Cushing, M. C., et al. 2009, ApJ, 702, 154
Stevenson, K. B., Harrington, J., Nymeyer, S., et al. 2010, Nature, 464, 1161
Struve, Otto, 1952, The Observatory, 72, 199
Swain, M. R., Vasisht, G., Tinetti, G., et al. 2009, ApJ, 690, L114
Tinetti, G., Vidal-Madjar, A., Liang, M.-C., et al. 2007, Nature, 448, 169
Traub, W., Shaklan, S. & Lawson, P. 2007, In the Spirit of Bernard Lyot, ed. P. Kalas
Volk, K., Blum, R., Walker, G., & Puxley, P. 2003, IAU Circ., 8188, 2
von Braun, K., Kane, S. R., & Ciardi, D. R. et al. 2009, ApJ, 702, 779
Wordsworth, R., Forget, F., Selsis, F., et al. 2011, ApJ, 733L, 48